\documentclass[conference]{IEEEtran}
\IEEEoverridecommandlockouts
\usepackage{cite}
\usepackage{xspace} 
\usepackage{amsmath,amssymb,amsfonts}
\usepackage{algpseudocode}
\usepackage{graphicx}
\usepackage{textcomp}
\usepackage{xcolor}
\usepackage{hyperref}

\def\BibTeX{{\rm B\kern-.05em{\sc i\kern-.025em b}\kern-.08em
    T\kern-.1667em\lower.7ex\hbox{E}\kern-.125emX}}

\hypersetup{
    colorlinks=true,
    linkcolor=black,
    filecolor=black,      
    urlcolor=blue,
    citecolor=black,
}
\newif\ifarxiv
\arxivtrue

\usepackage[utf8]{inputenc}
\usepackage[english]{babel}

\ifarxiv
\else
    \usepackage{minted}
    \usepackage{listings}
    \definecolor{keywordcolor}{rgb}{0.13,0.13,1}
    \definecolor{greenComments}{RGB}{63,127,95}
    \lstdefinelanguage{PSEUDO}{
        keywordstyle=[2]\color{blue},
        keywords=[2]{algorithm, is, input, output, for, each, in, do, while, return, if, then, else, end, and, or},
        sensitive=false,
        morestring=[b]",
        morecomment=[l]{//},
        }
    
    \lstdefinestyle{pseudo-style}{
        language=PSEUDO,
        backgroundcolor=\color{white},   
        keywordstyle=\color{blue},
        basicstyle=\ttfamily\scriptsize,
        breakatwhitespace=false,         
        breaklines=true,                 
        captionpos=b,                    
        keepspaces=true,                 
        showspaces=false,                
        showstringspaces=false,
        showtabs=false,                  
        tabsize=2,
        columns=fullflexible,
    }
    
    \lstnewenvironment{pseudo-listing}[1][]{
        \lstset{style=pseudo-style, caption=#1}
      }
      {}
\fi

\begin{document}

\def \midd {NLSC}
\def \middleware {\emph{\midd}\xspace}
\def \middlewarebf {\emph{\textbf{\midd}}\xspace}

\title{NLSC: Unrestricted Natural Language-based Service Composition through Sentence Embeddings}

\author{\IEEEauthorblockN{Oscar J. Romero}
\IEEEauthorblockA{\textit{Machine Learning Department} \\
\textit{Carnegie Mellon University}\\
5000 Forbes Av., Pittsburgh}
\and
\IEEEauthorblockN{Ankit Dangi}
\IEEEauthorblockA{\textit{Machine Learning Department} \\
\textit{Carnegie Mellon University}\\
5000 Forbes Av., Pittsburgh}
\and
\IEEEauthorblockN{Sushma A. Akoju}
\IEEEauthorblockA{\textit{Machine Learning Department} \\
\textit{Carnegie Mellon University}\\
5000 Forbes Av., Pittsburgh}
}

\maketitle

\begin{abstract}
Current approaches for service composition (assemblies of atomic services) require developers to use: (a) domain-specific semantics to formalize services that restrict the vocabulary for their descriptions, and (b) translation mechanisms for service retrieval to convert unstructured user requests to strongly-typed semantic representations. In our work, we claim that the effort to developing service descriptions, request translations, and service matching could be reduced using unrestricted natural language; allowing both: (1) end-users to intuitively express their needs using natural language, and (2) service developers to develop services without relying on syntactic/semantic description languages. Although there are some natural language-based service composition approaches, they restrict service retrieval to syntactic/semantic matching. With recent developments in Machine learning and Natural Language Processing, we motivate the use of Sentence Embeddings by leveraging richer semantic representations of sentences for service description, matching and retrieval. Experimental results show that service composition development effort may be reduced by more than 36\% while keeping a high precision/recall when matching high-level user requests with low-level service method invocations. 
\end{abstract}


\vspace{-0.2cm}
\section{Introduction and Related Work}
\vspace{-0.2cm}

\emph{Service} is any software component, data, or hardware resource on a device that is accessible by others~\cite{Chakraborty:2005}. \emph{Service composition} is the process of aggregating such reusable atomic services to create complex compositions. 
Existing research can be seen in two directions, where, (a) both atomic and composite services are defined using description languages (such as BPEL4WS, OWL-S, and WSDM) in terms of service input/output, pre- and post-conditions, fault handling and invocation mechanisms. Such service descriptions serve as inputs to orchestration engines~\cite{Sirin:2004,Tari:2010} that generate declarative specification of workflows to compose different services; and (b) architectural middlewares~\cite{Chakraborty:2005,Tomazini:2017,Hadj:2017} that assume a declarative specification of a composition. 
A substantial amount of effort is required in both directions to define and integrate services, mainly due to: (a) the use of domain-specific languages and semantics for service descriptions and compositions; (b) strongly-typed orchestration languages (e.g., BPEL, WSDL, OWL-S, etc.) restricting heterogeneous service composition; (c) statically specified compositions that create design-time couplings preventing dynamic adaptation; and (d) there is more than one composite service description languages: different ontologies have been designed resulting in different vocabularies, thwarting true semantic interoperability, so technologies have yet to converge and standardize~\cite{Stavropoulos:2013}. 
In natural language-based service composition middleware, end-users interact instinctively with systems in natural language and expect the system to identify services that meet their goals. These kind of middleware can be broadly categorized as those that: (a) apply restrictions on how the user expresses the goal with sentence templates and then use structured parsing to match against service descriptions \cite{DBLP:journals/jsw/BoscaCVM06,6384197}; (b) construct semantic graphs to represent service descriptions and match against a lexical database such as WordNet to compute concept similarity \cite{1546104,4801832,5283917}; and (c) match partially-observable natural language request with semantics of service description expressed using semantic web services (OWL-S, VDL) \cite{10.1007/978-3-642-17358-5_54,CHARIF200633}. Limitations with these approaches include: (a) complex linguistic processing that requires additional natural language processing (NLP) techniques: structured parsing, extracting parts-of-speech, stop-word removal, spell-checking, stemming, and text segmentation; (b) inclusion of lexical databases such as WordNet or domain-specific ontologies; and (c) a weaker concept representation and similarity score for semantic matching that does not account for sentence context. 

\textbf{\emph{Research Questions:}} to overcome the above, we address: 
\noindent\emph{RQ1: How to reduce the amount of development effort and complexity to develop service compositions?} \newline
\noindent\emph{RQ2: How can both end-users and developers create service compositions in an intuitive, efficient, and dynamic way using natural language-based descriptions?}
%

\textbf{\emph{Main Contributions:}} we address RQ1 by removing effort-consuming engineering practices, such as: (a) formal service descriptions that use syntactic/semantic representations; and (b) orchestration processes that use domain-specific languages. Additionally, we provide an automated OSGi-based toolchain for service modularity, service discovery, service deployment, and service execution. Our toolchain allows transparently deploying OSGi components to either cloud-based applications or mobile Android-based apps.  
And we addresses RQ2 by developing \middleware, a Natural Language-based Service Composition Middleware that: (a) allows users to express template-free service requests using natural language without complex linguistic processing; (b) avoids the need for lexical databases, semantic graphs, and domain-specific ontologies; (c) generates dynamic service compositions by directly binding  high-level user requests to low-level service invocations without having to define ontological service descriptions or strongly-typed well-defined interfaces; and (d) uses a stronger representation of sentence semantics to characterize words and concepts that account for word usage in context to the sentence by applying a state-of-the-art pre-trained semantic representation model of English language.  
The remaining of this paper is organized as follows: Section~\ref{sec_background} presents the background and  motivating example. Section~\ref{sec_approach} details the design and implementation and Section~\ref{sec_results} reports the experimental results. We introduce the related work and conclude the paper in Section~\ref{sec_related_work} and Section~\ref{sec_conclusions}, respectively.
%

\section{Background and Motivation}
\label{sec_background}

\subsection{Service Composition Middleware Model} 
According to the Service Composition Middleware (SCM) model~\cite{Ibrahim:2009} (a high-level abstraction model that does not consider a particular service technology, language, platform or algorithm used in the composition process), middleware for service composition can be largely classified into four main modules as follows: \emph{Translation, Generation, Evaluation}, and \emph{Execution}. 
In SCM, applications may send requests to middleware using diverse specification languages or techniques, and the \emph{Translator} converts these request descriptions into a system comprehensible language (i.e., formal languages and models) that can be used by the middleware. 
%
Once translated, the request specification is sent to the Generator, which provides the needed functionality by composing the available services, and generating one or several composition plans. This service composition is technically performed by chaining interfaces using either a syntactic or semantic method matching (or both). 
Then, the Evaluator chooses the most suitable composition plan depending strongly on many criteria like application context, the service technology model, the non functional service QoS (Quality of Service) properties, etc. 
Finally, the Builder executes the selected composition plan and produces an implementation corresponding to the required composite service. Once the composite service is available, it can be executed by the application that required its functionality. 

\subsection{Motivational Example} 
\begin{figure}[tbp]
\vspace{-0.3cm}
\centerline{\includegraphics[width=\columnwidth]{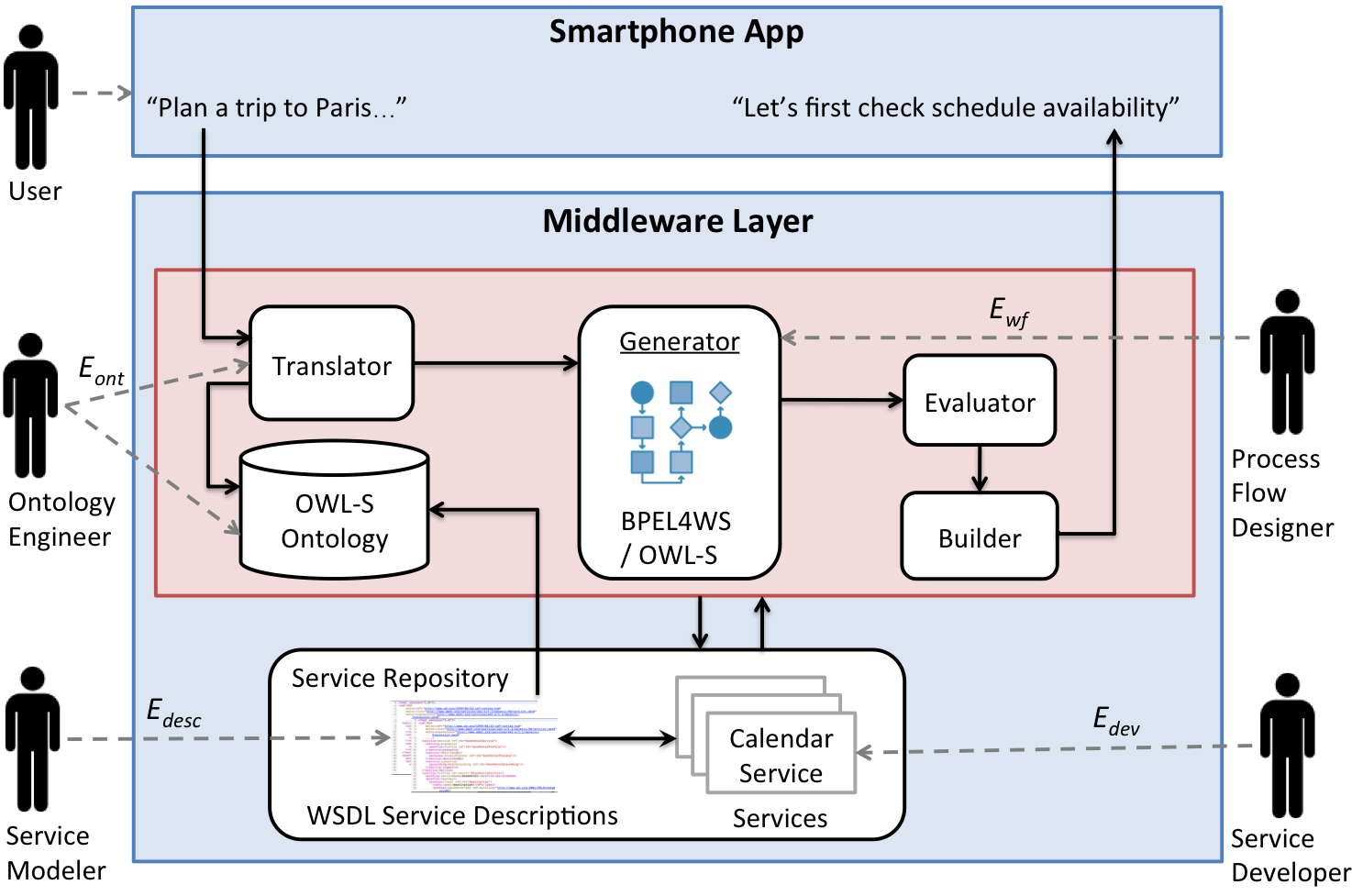}}
\vspace{-0.5cm}
\caption{Motivating Example: Plan a trip scenario}
\label{fig_motivation}
\vspace{-0.4cm}
\end{figure}

Suppose the user is planning a trip to Paris on a specific range of dates (main goal) using a smartphone that does not have a \emph{trip planner} service or app installed. This main goal can be decomposed into sub-goals such as: (1) check schedule availability on dates, (2) look for flights cheaper than $\$$700, (3) book the chosen flight, (4) search for hotels under $\$$100/night near downtown, (5) book the selected hotel, (6) check the weather conditions for given dates, (7) if weather conditions are bad, look for indoor activities to do, (8) otherwise, look for outdoor activities to do. To address this scenario (see Figure~\ref{fig_motivation}), a \emph{service developer} would create atomic service interfaces and implementations for services such as Maps, Calendar, FlightBooking etc.; a \emph{service modeler} would define the service interface contracts using WSDL; an \emph{ontology engineer} would maintain the trip-planning ontologies and ensure consistency with OWL-S models; and a \emph{process flow designer} would investigate explicit declarative alternatives to generate a service composition that addresses user's goal.

\subsection{Problem Statement} 
In the example above, the total effort required is the sum of partial efforts. Let $E_{T} = \sum_{i=1}^{n} E_{i}$, where, $E_{i} \in \{E_{dev}, E_{desc}, E_{ont}, E_{wf}\}$ such that $E_{dev}$, $E_{desc}$, $E_{ont}$, $E_{wf}$ are effort amounts to develop a service implementation, generate a WSDL service description, create/maintain an OWL-S ontology, and maintain a BPEL4WS workflow, respectively (for simplicity, we ignore additional efforts for testing, CI/CD, etc.) These efforts increase exponentially when service requirements change continuously, there are inconsistencies on service contracts, or developers don't have the proper skillset.

\subsection{Goals} 
Our scientific intuition leads us to hypothesize that a data-driven approach (using large text corpus and datasets of common-sense sentences) not only could minimize the effort of developing service compositions by removing the need of specifying strongly-typed, syntactically/semantically well-defined, domain-dependent service descriptions, but also could outperform traditional semantic-driven approaches that require continuous validation of consistency due to human designers' biased models.
Therefore, driven by our RQs, our goal is two-fold: (a) to reduce the total effort of integrating new services into a composition by merging/replacing some of the development tasks previously described without affecting either system performance or the quality of service compositions; and (b) to automatically bind unrestricted natural language user requests to unstructured natural language service descriptions with control structures for composition. 

\subsection{OSGi}
OSGi (Open Services Gateway initiative) technology~\cite{OSGi} is a set of specifications that define a dynamic component system for Java. These specifications enable a development model where an application is composed of several components that are packaged as bundles. Components communicate locally and across the network through services. Services have an API that is defined in a Java package. Some of the most known OSGi-based middleware for service composition are:~\cite{Ibrahim:2009.1,Choonhwa:2007,Pourreza:2006,Redondo:2007}. We use OSGi as a backbone for connecting multiple service implementations, providing a mean for the exchange of information between them. 

%
\section{Approach}
\vspace{-0.1cm}
\label{sec_approach}

\subsection{Preliminaries} 
As it is generally considered in the literature~\cite{Balzer:2004,Tari:2010}, we distinguish two types of services~\cite{Balzer:2004,Tari:2010}: \emph{abstract} and \emph{concrete} services. Formally, a concrete service $cs_{i}$ is a tuple  $\langle cs_{i}^{in}, cs_{i}^{out}, cs_{i}^{prec}, cs_{i}^{postc}, cs_{i}^{QoS} \rangle$ that performs a functionality by acting on input data ($cs_{i}^{in}$) to produce output data ($cs_{i}^{out}$), with pre-conditions ($cs_{i}^{prec}$), post-conditions ($cs_{i}^{postc}$) and Quality of Service ($cs_{i}^{QoS}$) requirements. An abstract service $as_i$ is a tuple $\langle as_{i}^{in}, as_i^{out}, as_i^{cs} \rangle$ realized by several concrete services $as_i^{cs} \in \{cs_{(i,1)}, cs_{(i,2)},..., cs_{(i,n)}\}$ that offer the same functionality with input parameters ($as_{i}^{in}$), output parameters ($as_{i}^{out}$) such that $\forall cs_{(i,j)}, cs_{(i,k)} \in as_{i}^{cs} / (as_{i}^{in} = cs_{(i,j)}^{in} \cap cs_{(i,k)}^{in}) \wedge (as_{i}^{out} = cs_{(i,j)}^{out} \cap cs_{(i,k)}^{out})$.

\vspace{-0.cm}
\subsection{Reference Architecture} 
\begin{figure}[t]
\centerline{\includegraphics[width=\columnwidth]{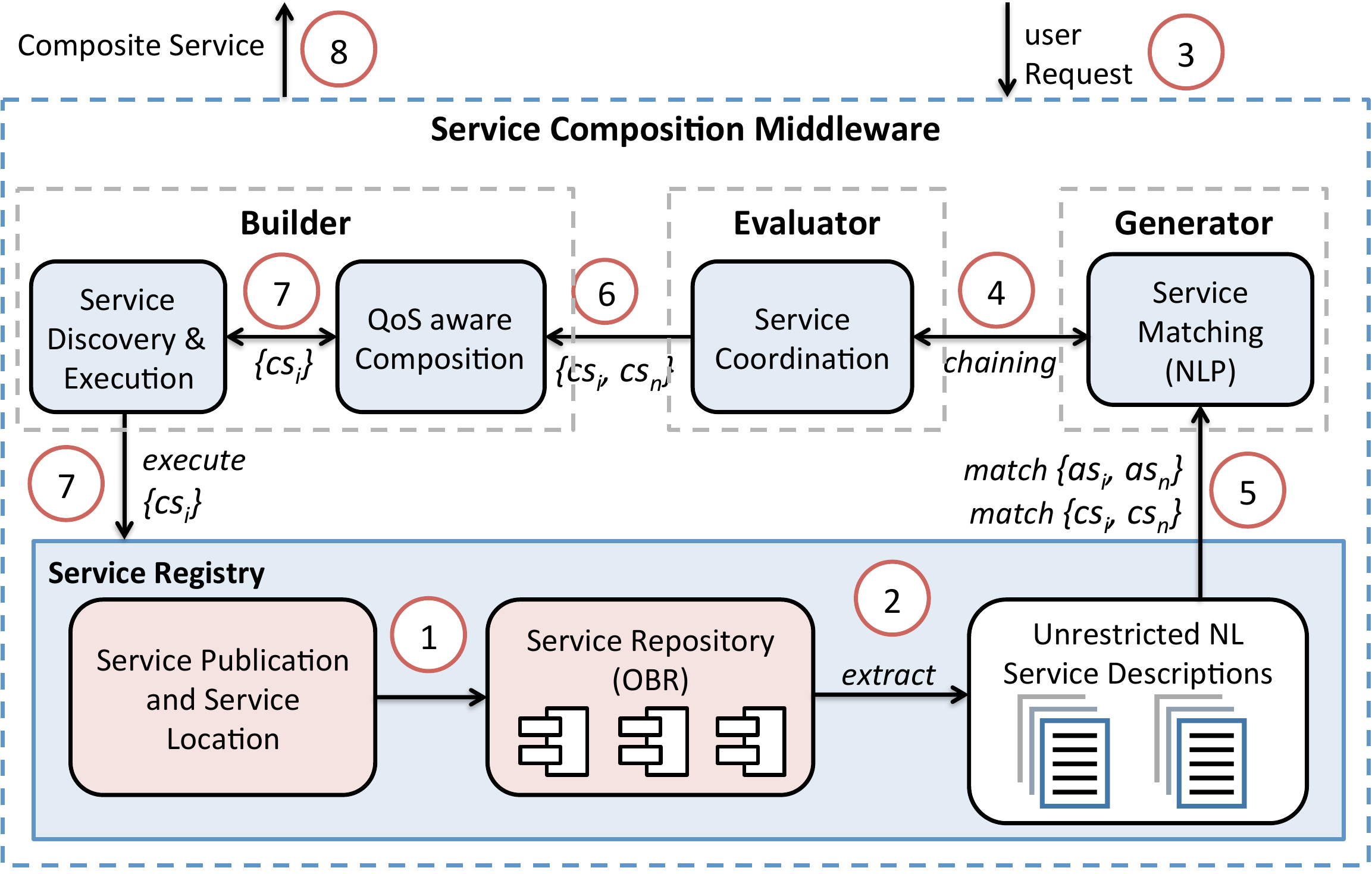}} \vspace{-0.3cm}
\caption{Reference Architecture for a NLSCM}
\vspace{-0.4cm}
\label{fig_architecture}
\end{figure}

Figure~\ref{fig_architecture} presents a reference architecture for service composition that will help highlight where our contributions lie. 
\emph{Step 1:} service developers continuously implement, integrate, deploy and publish service components (either abstract or concrete). Developers add unstructured and unrestricted natural language descriptions (in the form of plain code annotations) to each single service component and its atomic methods. In comparison, traditional approaches would include additional steps (i.e., service description using WSDL, creation and validation of OWL-S ontology, etc.). \emph{Step 2:} an automated process extracts those descriptions from the code annotations and puts them on a separate repository. \emph{Step 3:} the end-user, an application developer, or a top-tier application makes a service request (e.g., \emph{``I want to plan a trip to Paris from Sept. 29 to Oct. 11''}). \emph{Step 4:} a coordination system is in charge of orchestrating the high-level assembly of abstract services by chaining service pre- and post-conditions and matching data types (traditional approaches would include additional steps for creating complex graph-based, workflow-based or rule-based plans). \emph{Step 5:} the service matching is performed using two NLP techniques: Sentence embeddings and Named-Entity Recognition, and returns a set of abstract services and their corresponding concrete service candidates. Steps 4 and 5 are repeated until a composition plan is tailored. \emph{Step 6:} a mechanism validates the QoS requirements by selecting a sub-set of candidate concrete services that are to be executed. \emph{Step 7:} using a service discovery mechanism, the sub-set of concrete service candidates are looked up in the registry and service availability for those is confirmed. \emph{Step 8:} lastly, a composite service is generated and executed. 
Compared to the SCM model, we suppress the Translator module and only keep the remaining ones (Generator, Evaluator, and Builder). We overcome the need for the Translator as our approach does not use intermediate representations such as ontologies, graph-based models, and so forth, rather, we provide a direct correlation between the request and the service description which are both expressed using unrestricted natural language.

\subsection{Service Development Process}
\label{sec_toolchain}

\begin{figure}[htbp]
\centerline{\includegraphics[width=\columnwidth]{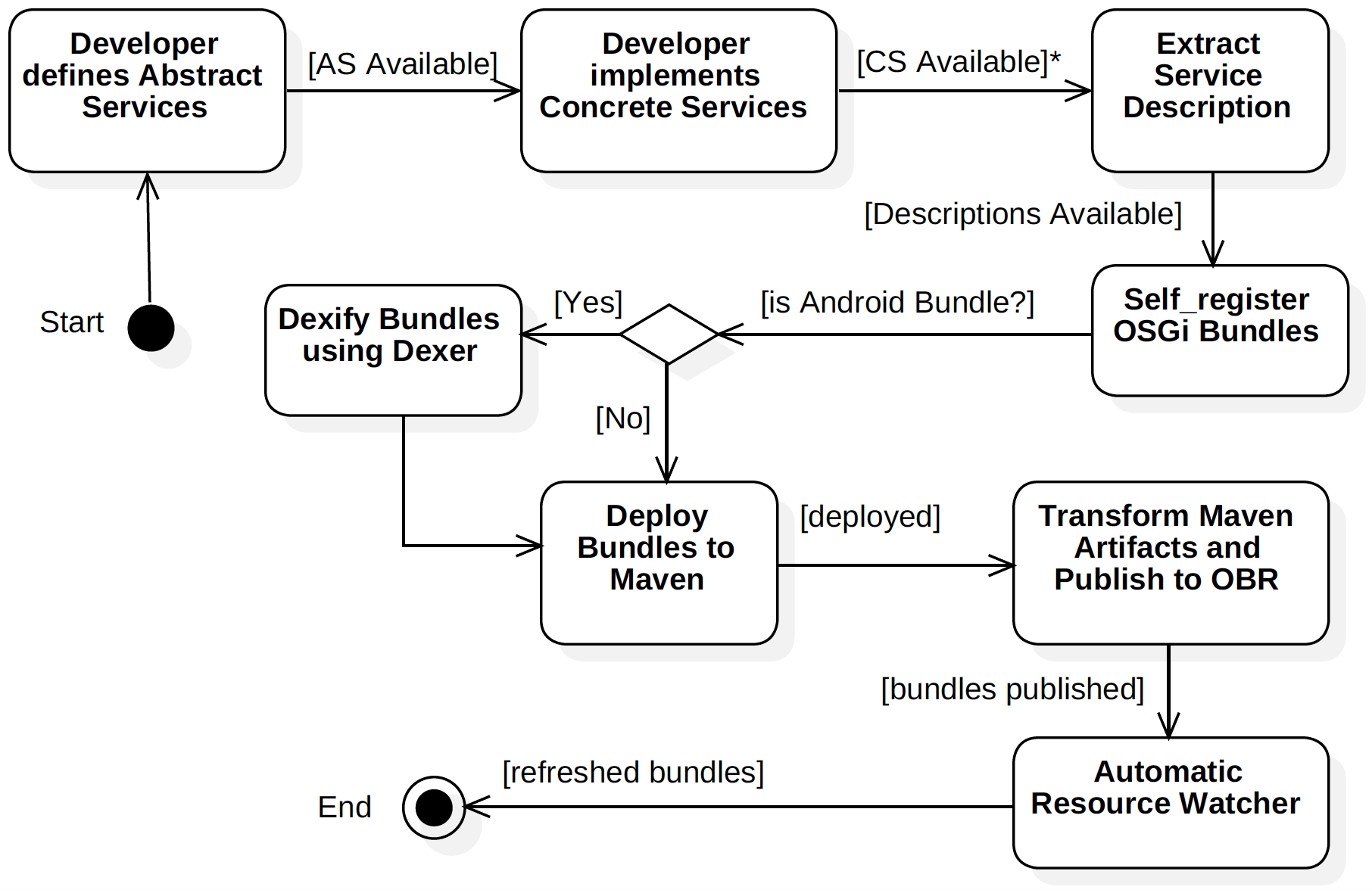}} 
\vspace{-0.3cm}
\caption{Workflow for Service Development using \middleware}
\vspace{-0.3cm}
\label{fig_workflow}
\end{figure}

\middleware was architected to be deployed on distributed environments and to support different kind of client applications (e.g., standalone, web, mobile, etc.). This requirement, along with dynamicity, low-latency, high-performance, modularity, and support to Android Runtime Environment (ART), were the main architectural significant requirements we took into consideration over its implementation~\cite{Romero:2018}. Extending our initial definition of ``service'', we consider that an Android-based service can be a device sensor, a local service (installed on the Android device), or an app, hence, we avoid making the assumption that only a web service model should be applicable to \middleware. Given these requirements, a suitable solution for developing dynamic service components for Android is the OSGi technology~\cite{Kalinowski:2015}. 
Most of the OSGi-Android approaches~\cite{Bouzefrane:2011,Chang:2010} are based on Apache Felix~\cite{Felix:2015}, an implementation of the OSGi Framework and Service platform. 
%
%
In \middleware, composite services are created from a dynamic assembly of black box components, executing in a local Felix container, which does not provide mandatory non-functional services. Services do not contain any reference between them at design-time, and respect black box and late-binding concepts. We tried to keep the intervention of application developers to the minimum, automating as much as possible the discovery, composition, invocation and interoperability of services and, therefore, reducing the development effort. To that purpose, we developed a set of tools to simplify the process of service development, and promote agile development and continuous integration into \middleware. This process is shown in Figure~\ref{fig_workflow} and described next.

\subsubsection{Abstract Services Description}
\label{sec_descriptions}

Common OSGi-based approaches for service composition provide modularity, though still a Translator is required to guarantee interoperability between semantic and syntactic service description languages that are both heterogeneous. Prior work shows the heavy cost of the syntactic and semantic matching~\cite{Ibrahim:2009}. In our work, we replace effort-consuming syntactic/semantic service descriptions (WSDL, OWL-S, etc.) by intuitive code annotations that allow developers to add unrestricted natural language descriptions to each service component and its methods. More specifically, we provide 2 code annotations: \emph{@Description} annotation allows developers to add a set of possible capabilities for an atomic service method (in terms of what the service method can do), and \emph{@ArgDesc} let developers add descriptions to method arguments that can be later used for argument type disambiguation. Alternatively, \middleware can also load these annotations from plain files in order to decouple them from the programming language.

\ifarxiv
   \begin{figure}[htbp]
    \centerline{\includegraphics[width=\columnwidth]{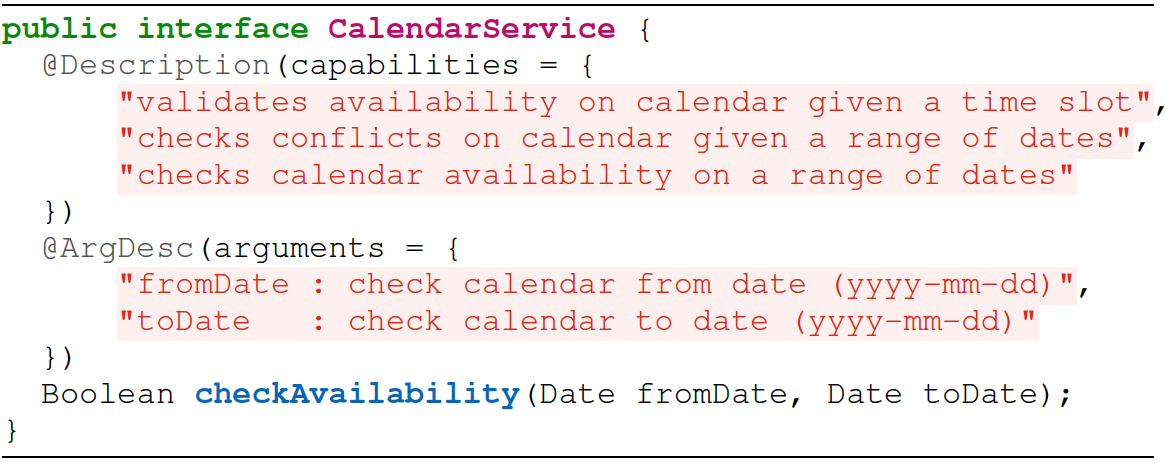}} 
    \vspace{-0.3cm}
    \caption{Abstract service description for CalendarService}
    \vspace{-0.1cm}
    \label{lst_check}
    \end{figure}
\else

\begin{listing}[ht]
\begin{minted}
[
frame=lines,
framesep=1mm,
baselinestretch=1,
fontsize=\scriptsize
]{java}
public interface CalendarService {
  @Description(capabilities = {
      "validates availability on calendar given a time slot",
      "checks conflicts on calendar given a range of dates",
      "checks calendar availability on a range of dates"
  })
  @ArgDesc(arguments = {
      "fromDate : check calendar from date (yyyy-mm-dd)",
      "toDate   : check calendar to date (yyyy-mm-dd)"
  })
  Boolean checkAvailability(Date fromDate, Date toDate);
}
\end{minted}
\caption{Abstract service description for CalendarService}
\label{lst_check}
\end{listing}    
\fi
From \ifarxiv Figure \else Listing \fi \ref{lst_check}, it is worth noting that: a) given an abstract service $as_i$ exposes a set of methods $m_i = \{m_{(i,1)}...m_{(i,n)}\}$, where, a method is described as a tuple $\langle cap_{m_{i}}, args_{m_{i}}, argd_{m_{i}} \rangle$ where $cap_{m_{i}}$ is an arbitrary number of capabilities such that $|cap_{m_{i}}| \geq 1$, $args_{m_{i}}$ is a set of method arguments, and $argd_{m_{i}}$ is a set of argument descriptions corresponding to $args_{m_{i}}$, such that $|args_{m_{i}}| = |argd_{m_{i}}| \geq 0$; b) the list of arguments in @ArgDesc maps each description onto a method argument using the name of the argument, a description in natural language, and (optional) the format or type of the argument (used to disambiguate with the user or when using the Named-entity Recognition technique). c) developer does not need to do extra effort defining a service description or ontology using WSDL, OWL-S, etc. (especially when developers are unfamiliar with such languages, but even if they are, using unrestricted natural language descriptions is more intuitive and easy-to-deploy).

\subsubsection{Concrete Services Implementation}
A concrete service $cs_i$ inherits method descriptions from abstract service $as_i$. It defines non-functional, platform-specific QoS requirements for methods to guarantee service execution if and only if they are met. For illustration purposes, let's continue with our motivating example and assume the platform is Android. An abstract service could be \emph{CalendarService} ($as_{cal}$) whereas concrete services could be \emph{GoogleCalendarService} ($cs_{gc}$) and \emph{YahooCalendarService} ($cs_{yc}$). \middleware provides a set of pre-defined QoS annotations for Android, though they can extended: @BatteryQoS is a categorical value for battery level consumption that indicates whether the service is battery-intensive (e.g., LOW\_BATTERY, HALF\_CHARGED, FULLY\_CHARGED), @ConnectivityQoS is a categorical value that determines whether the service requires deveice's wifi connection or if it runs locally, etc.  
\ifarxiv
    \begin{figure}[htbp]
    \vspace{-0.3cm}
    \centerline{\includegraphics[width=\columnwidth]{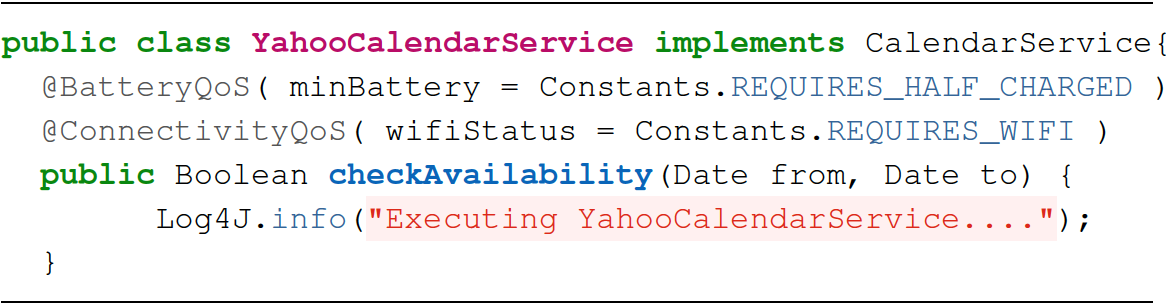}} 
    \vspace{-0.3cm}
    \caption{QoS-awareness for YahooCalendarService}
    \vspace{-0.3cm}
    \label{lst_yahoo}
    \end{figure}
\else
\begin{listing}[ht]
\begin{minted}
[
frame=lines,
framesep=2mm,
baselinestretch=1.2,
fontsize=\scriptsize
]{java}
public class YahooCalendarService implements CalendarService{
  @BatteryQoS( minBattery = Constants.REQUIRES_HALF_CHARGED )
  @ConnectivityQoS( wifiStatus = Constants.REQUIRES_WIFI )
  public Boolean checkAvailability(Date from, Date to) {
        Log4J.info("Executing YahooCalendarService....");
  }
\end{minted}
\caption{QoS-awareness for YahooCalendarService}
\label{lst_yahoo}
\end{listing}
\fi
From \ifarxiv Figure \else Listing \fi \ref{lst_yahoo}, we observe that YahooCalendarService will be executed only if its QoS features are met, that is, the smartphone's battery has to be at least half charged and it should be connected to the WiFi, otherwise, another concrete service that implements CalendarService is discovered.

\subsubsection{Service Descriptions Extractor}
\label{sec_extractor}
Using Java reflection, this tool automatically generates a plain file with all service method descriptions which is further used by the Service Matching module (Section \ref{sec_matching}). Additionally, it generates a metadata file with method argument descriptions and QoS values that are used at the time of service execution after services are grounded (Section \ref{sec_execution}).

\subsubsection{OSGi bundle self-registration}
Both abstract and concrete services are deployed as OSGi bundles. This tool automatically generates an implementation of a BundleActivator (an OSGi interface that manages bundle's lifecycle) and injects code on the \emph{start()} and \emph{stop()} methods to self-register or self-unregister the bundle against the Felix Framework. 

\subsubsection{Dexifying bundles}
Android Runtime does not use Java bytecode, instead, Android programs are compiled into .dex (Dalvik Executable) files. Thus, we developed Dexer, a tool that automatically transforms the Java class files compiled by a regular Java compiler into a class file format that can be executed on the Android runtime. In other words, Dexer automatically converts an OSGi bundle into an executable Jar that can be later executed on the Android platform. 

\subsubsection{OSGi Maven Deployer}
Transforms application Jars to OSGi bundles that are then automatically deployed to a remote Maven repository, which makes the artifacts accessible to application developers and service runtime environment. 

\subsubsection{TAMO}
This is a tool that automatically transforms artifacts from a Maven repository (that holds OSGi bundle artifacts) to an OSGi Bundle Repository (OBR). Felix OBR provides a service that can automatically install a bundle, with its deployment dependencies, from a bundle repository, enabling location and discovery of the participating services during the composition process.

\subsubsection{ARW}
The Automatic Resource Watcher (ARW) pulls data periodically from an OBR in order to find new available services or updates for existing services. This functionality is critical for the service discovery phase during service execution because it allows re-configuration of services and enables the generation of compositions on-demand.

\vspace{-0.1cm}
\subsection{Service Matching/Selection}
\vspace{-0.1cm}
\label{sec_matching}
Current approaches on service composition perform service matching by doing syntactic and semantic interface matching, then the service evaluation is performed upon the input/output matching correctness. As we described before, semantic matching though useful is expensive in terms of effort (i.e., the construction of one single service involves the design, maintenance and consistency validation of syntactic/semantic representations carried out by ontology engineers and service designers) and computing time (the larger an ontology is, the longer it takes to perform semantic inference or concept graph search). Instead of using syntactic or semantic matching through the use of ontologies, we propose semantic service matching through the use of \emph{Sentence Embeddings}. 
In linguistics, and more specifically in feature learning techniques in natural language processing (NLP), both word embeddings and sentence embeddings are studied by the area of distributional semantics. Embeddings aim to quantify and categorize semantic similarities between linguistic items based on their distributional properties in large samples of language data. Word embeddings capture the idea that is possible to express ``meaning'' of words using a vector, so that the cosine of the angle between the vectors captures semantic similarity (``cosine similarity'' property). Sentence embeddings and text embeddings extend word embeddings to sentences and paragraphs: they use a fixed-dimensional vector to represent a short piece of text, e.g., a sentence or a small paragraph. 
Sentence embeddings account for sentence context using the words in the sentence (based on the distributional hypothesis, that is, sentences that occur in the same contexts tend to have similar meanings), providing a richer semantic representation that makes it a reasonable choice for using natural-language descriptions for service matching rather than using other existing natural language-based service matching approaches where only word-level matching is performed ignoring the context. 
We used \emph{sent2vec}\cite{Pagliardini:2018} to perform text understanding using sentence embeddings. sent2vec is a model that can be seen as an extension of the CBOW (Continuous Bag of Words\cite{Mikolov:2013}) where the training objective is to train sentences instead of word embeddings. sent2vec has demonstrated that the empirical performance of the resulting general-purpose sentence embeddings significantly exceeds the state of the art, while keeping the model simplicity as well as training and inference complexity exactly as low as in averaging methods. 

\begin{figure}[t]
\centerline{\includegraphics[width=\columnwidth]{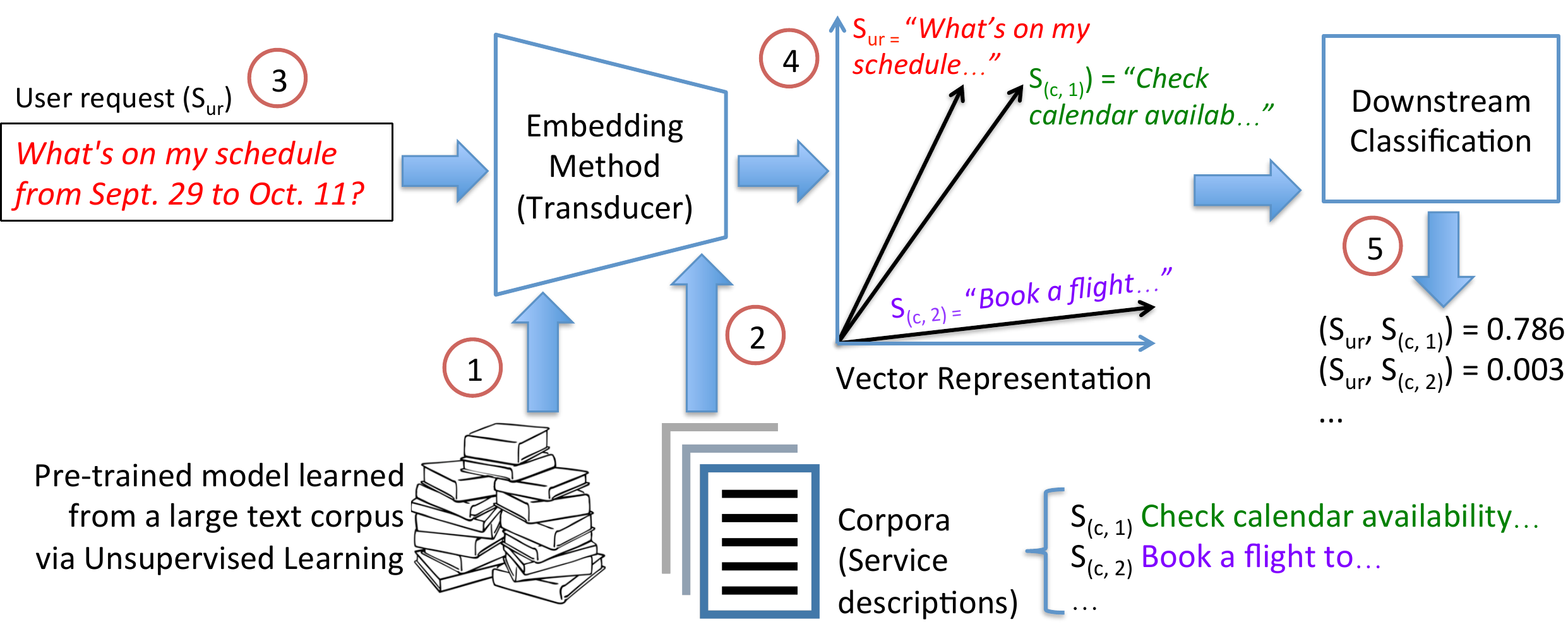}} 
\vspace{-0.2cm}
\caption{Pipeline for the Service Matching using Text Embedding}
\vspace{-0.5cm}
\label{fig_pipeline}
\end{figure}

Let us revisit our motivating example, where user's goal is \emph{``plan a trip to Paris on...''}. The required pipeline to carry out service matching is shown in Figure~\ref{fig_pipeline}. 
In step 1, a pre-trained model is learned through unsupervised machine learning over a large dataset of sentences (in our experiments we used two datasets, one with 19.7 billion sentences from Tweeter entries, and another with 1.9 billion sentences from Wikipedia entries) using sent2vec training mode. As a result of this step, sentences and their meaning are mapped onto vectors of real numbers (embeddings). 
In step 2, a textual corpus $C$ is automatically generated by \middleware by extracting the service descriptions from the annotations that developers add to their abstract services during the development phase (as explained in sections~\ref{sec_toolchain} and \ref{sec_extractor}), such that $C = \{as_1^{m} \cup as_2^{m}... \cup as_n^{m}\}$, where $m$ is the set of service method descriptions and $n$ is the total number of abstract services $as$. 
In step 3, user makes a service request $S_{ur}$. For instance, assume that user makes the following request as part of the required steps to achieve his/her goal of planning a trip to Paris: $S_{ur} =$ \emph{``check what's on my schedule from Sept. 29 to Oct. 11?''}. 
In step 4, the user's request $S_{ur}$ is fed as input into sent2vec and then mapped to a vector (embedding) in a n-dimensional space. 
Finally, in step 5, a downstream classification finds the nearest neighboring sentence feature (the optimal match for $S_{ur}$ in terms of a higher sentence embedding match) by computing the sentence similarity (correlation of the cosine similarity between two embeddings) for each pair of sentences ($S_{ur}, S_{(c, i)}$), where $S_{(c, i)}$ is the $i$-th service description contained by the corpus $C$. 
For instance in Figure~\ref{fig_pipeline}, after downstream classification, the sentence similarity between user request $S_{ur} =$ \emph{``what's on my schedule...''} and the service method description $S_{(c, 1)} =$ \emph{``check calendar availability...''} for method \emph{checkAvailability()} that belongs to the abstract service CalendarService will be higher (0.786), whereas the similarity between the same user request and service method description $S_{(c, 2)} =$ \emph{``book a flight...''} that belongs to FlightReservationService will be much lower (0.003). Given this example, the user request would be matched with method description checkAvailability on abstract service CalendarService. 
Service matching selects the most appropriate abstract service method based on a similarity-based selection as described in the pseudocode on \ifarxiv Figure \else Listing \fi ~\ref{lst_pseudo}. Basically, in order to be selected, an abstract service method must have the highest similarity, which should be above an upper threshold (t1 = 0.6); if not, abstract services that are in the range $(t2 \geq as \geq t1)$ and their similarities differ in less than delta (0.01) then they need to be disambiguated by the user; otherwise the similarity between user request and service method description is too low that no service can be selected and user needs to re-phrase the request. Values for thresholds and delta have been discovered empirically and have demonstrated satisfactory results.

\ifarxiv
    \begin{figure}[t]
    \centerline{\includegraphics[width=\columnwidth]{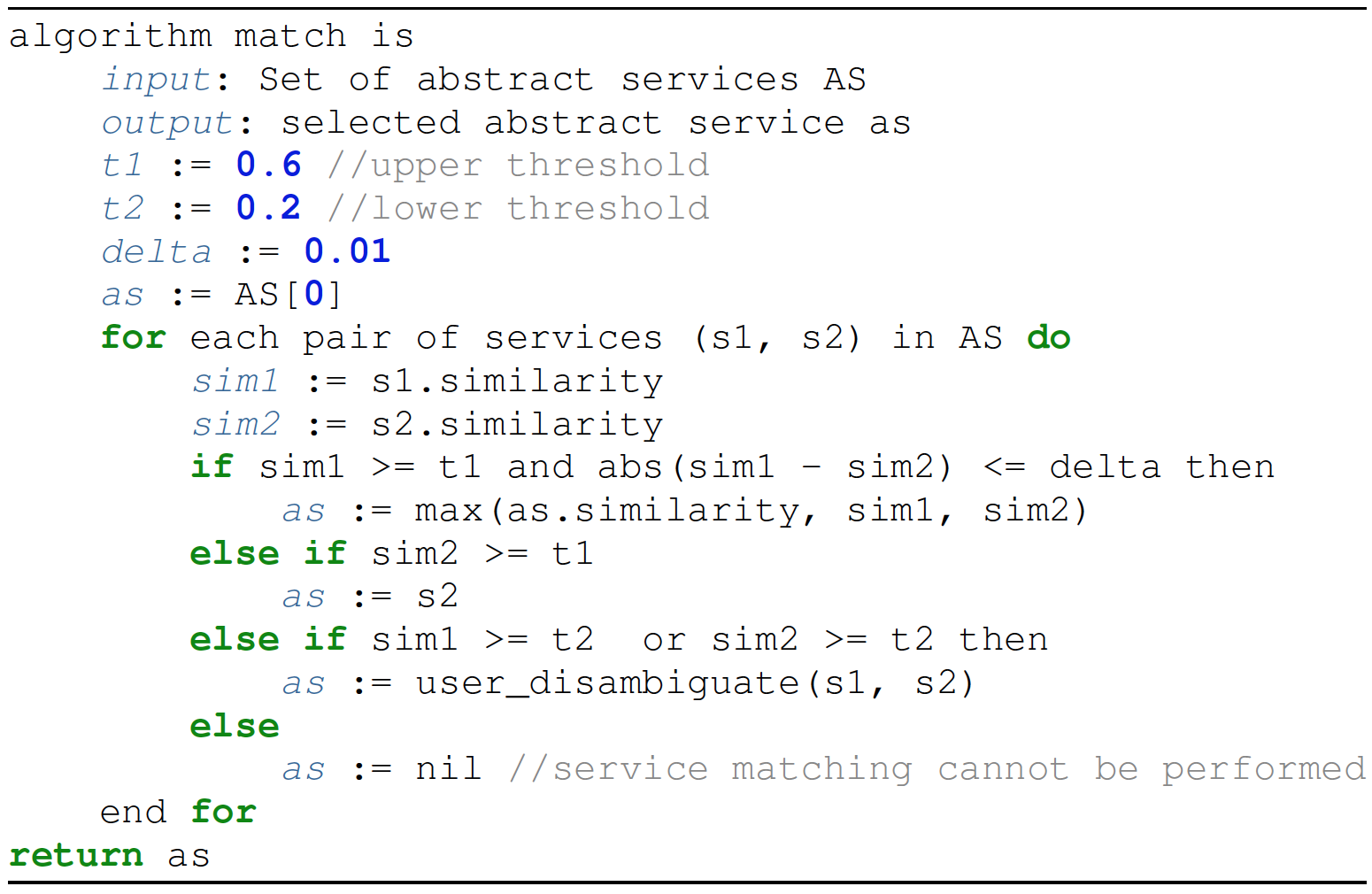}} 
    \vspace{-0.2cm}
    \caption{Pseudocode for Service Matching}
    \vspace{-0.5cm}
    \label{lst_pseudo}
    \end{figure}
\else
\begin{listing}[h]
\begin{minted}
[
frame=lines,
framesep=1mm,
baselinestretch=1,
fontsize=\scriptsize
]{c++}
algorithm match is
    input: Set of abstract services AS
    output: selected abstract service as
    t1 := 0.6 //upper threshold
    t2 := 0.2 //lower threshold
    delta := 0.01
    as := AS[0]
    for each pair of services (s1, s2) in AS do
        sim1 := s1.similarity
        sim2 := s2.similarity
        if sim1 >= t1 and abs(sim1 - sim2) <= delta then
            as := max(as.similarity, sim1, sim2)
        else if sim2 >= t1 
            as := s2
        else if sim1 >= t2  or sim2 >= t2 then
            as := user_disambiguate(s1, s2)
        else
            as := nil //service matching cannot be performed
    end for
return as
\end{minted}
\caption{Pseudocode for Service Matching}
\vspace{-0.1cm}
\label{lst_pseudo}
\end{listing}
\fi

\vspace{-0.2cm}
\subsection{Service Coordination}
\label{sec_coordination}
The Service coordination comprises three mechanisms: a short-term (working) memory where results are stored temporarily, data type disambiguation through entity matching and named-entity recognition, and a rule-based system that allows creating high-level assemblies of abstract services (composite services) by chaining pre/post-conditions. 

\textbf{\emph{Short-term Working Memory (WM):}}  the WM stores not only the partial results and inferences produced by the forward chaining process of the rule engine but also keeps updated information collected from sensors (in the case of an Android phone), user preferences, service status, and QoS features. WM is implemented as a Hash Table or Dictionary.

\textbf{\emph{Data type disambiguation:}} we disambiguate data types by using named-entity recognition. To that purpose, we use Stanford NER\cite{Finkel:2005}, a Java implementation that labels sequences of words in a text that are names of things, such as person and company names. It provides well-engineered feature extractors that annotates sentences with labels such as: NOUN, PERSON, COMPANY, NUMBER, MONEY, TIME, DATE, and LOCATION. However, since it provides a general implementation of (arbitrary order) linear chain Conditional Random Field (CRF) sequence models, it is possible to train customized models on labeled data extracted from service descriptions.
For example, suppose that the user request is: \emph{``look for flights to Paris for less than \$700''}. Then, Service Matching module (Section~\ref{sec_matching}) outputs a set of method descriptions and their corresponding abstract services along with a similarity score associated to each method description (based on the cosine similarity). Now, suppose that the abstract service method description shown in \ifarxiv Figure~\ref{lst_search} \else listing~\ref{lst_search} \fi is the best match for the user request. Once the abstract service method is selected, the type-based disambiguation is performed as follows: first, suppose that the value for argument ``from'' is provided by the WM (assuming that the system inferred this value from user's current location or extracted it from a previous user request, and then stored it into the WM). Now, user request provides two additional values, one string value (Paris) and one numeric value (700). Without further processing, the string value ``Paris'' could match either ``to'' or ``cabin'' string arguments, and the numeric value ``700'' could match either ``price'' or ``numPass'' numeric arguments. For this reason, we use NER to disambiguate the types for method searchFlights, that is, NER is able to infer that ``Paris'' is a LOCATION, ``\$700'' is MONEY, and ``flight'' is a NOUN. 

\ifarxiv
    \begin{figure}[t]
    \centerline{\includegraphics[width=\columnwidth]{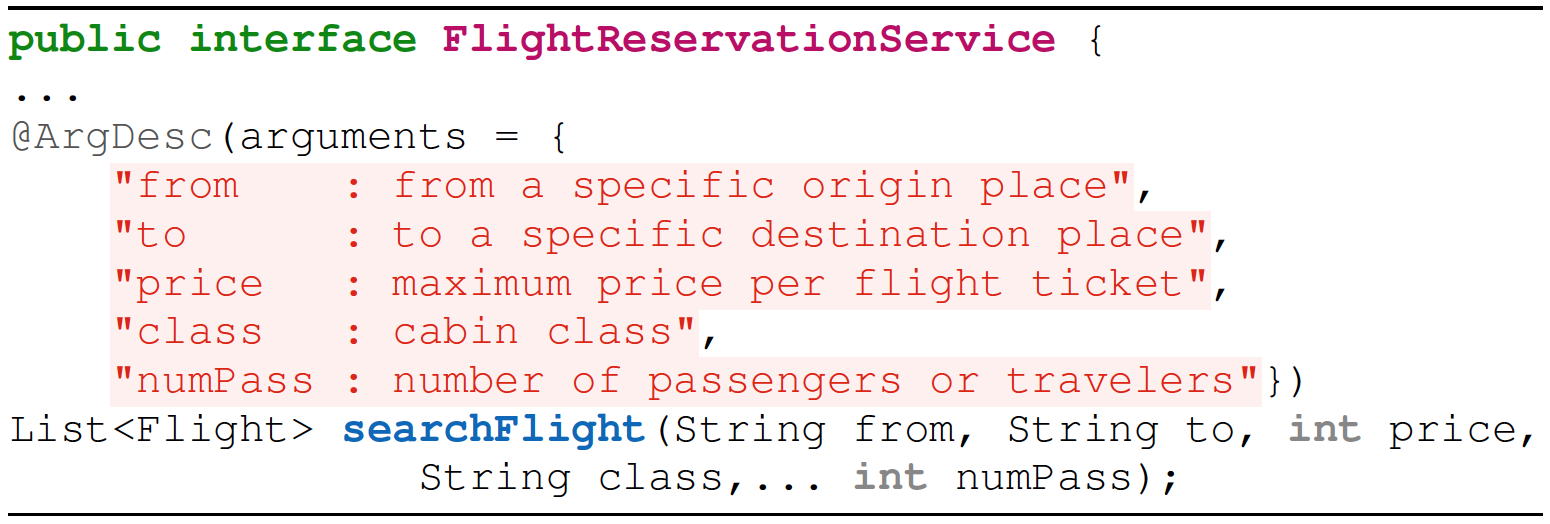}} 
    \vspace{-0.2cm}
    \caption{Argument description for searchFlight method}
    \vspace{-0.5cm}
    \label{lst_search}
    \end{figure}
\else
\begin{listing}[ht]
\begin{minted}
[
frame=lines,
framesep=1mm,
baselinestretch=1,
fontsize=\scriptsize
]{java}
public interface FlightReservationService {
...
@ArgDesc(arguments = {
    "from    : from a specific origin place",
    "to      : to a specific destination place",
    "price   : maximum price per flight ticket",
    "class   : cabin class",
    "numPass : number of passengers or travelers"})
List<Flight> searchFlight(String from, String to, int price, 
                String class,... int numPass);
\end{minted}
\caption{Argument description for searchFlight method}
\label{lst_search}
\end{listing}
\fi 
The next step is to automatically extract all the nouns from the argument descriptions of the service method (i.e., @ArgDesc annotations) using a Part-Of-Speech tagger such as Stanford POS Tagger~\cite{Toutanova:2003}. Now, using an Automatic Synonym Extractor (like WordNet synset or Word2Vec) over the resulting set of nouns from the previous step (e.g., origin, destination, place, price, flight, etc.) is possible to determine that the closest synonym for ``location'' (Paris) is ``place''. Since there are two places (origin and destination) but ``origin place'' was already resolved by the WM, then Paris is mapped onto argument ``to'' (destination place). Likewise, since the closest synonym for ``money'' is ``price'', then \$700 is mapped onto argument ``price''. Since there are no other information available, the remaining arguments (class, numPass, etc) are disambiguated directly with the user.

\textbf{\emph{Compositional Rules:}} the compositional rules allow linking different abstract service methods by chaining their pre- and post-conditions. For the sake of simplicity, we assume that service method pre-conditions correspond to service method arguments, while service method post-conditions correspond to service method returned elements. Rules can retrieve, add and remove information from the WM according to the current and future compositional needs. After executing a service method, the returned value (post-condition) is added to the WM, while during data type disambiguation and pre-condition matching values stored in WM are retrieved. Finally, WM contents can be removed by rules if they are not longer needed for the current composition. 
As a rule-based system to support the creation and execution of compositional rules we used \emph{easy-rules}\cite{EasyRules}, a lightweight yet powerful Java rule engine that can be executed in a wide variety of platforms, including Android. easy-rules also supports MVEL (MVFLEX Expression Language \cite{MVEL}), a hybrid dynamic/statically typed, embeddable Expression Language and runtime for the Java Platform. 
MVEL is typically used for exposing basic logic to end-users and programmers through weakly-typed (or non-typed) expressions. MVEL is dynamically typed (with optional typing), meaning that type qualification is not required in the source, which confers significant flexibility to our purpose of creating dynamic compositional rules based on unrestricted language descriptions. 
For instance, service methods ``searchFlight'' and ``bookFlight'' can be automatically chained using two MVEL rules as shown in \ifarxiv Figure \else Listing \fi ~\ref{lst_mvel}: when all pre-conditions of method ``searchFlight'' are stored into the WM (i.e., the condition part (when) of MVEL rule ``rule-search-flights'': flight.destination, flight.from, etc.) then an instance of the service FlightReservationService is obtained, then the method searchFlight is executed, and finally the results of the method execution are stored back into the WM (i.e., wm.put(`selectedFlights', ....)). Once the post-conditions of method searchFlight are stored into the WM, then the pre-conditions of method bookFlight are met so the second MVEL rule (rule-book-flight) can be triggered, and the process continues until no more information can be chained.
\ifarxiv
   \begin{figure}[htbp]
    \centerline{\includegraphics[width=\columnwidth]{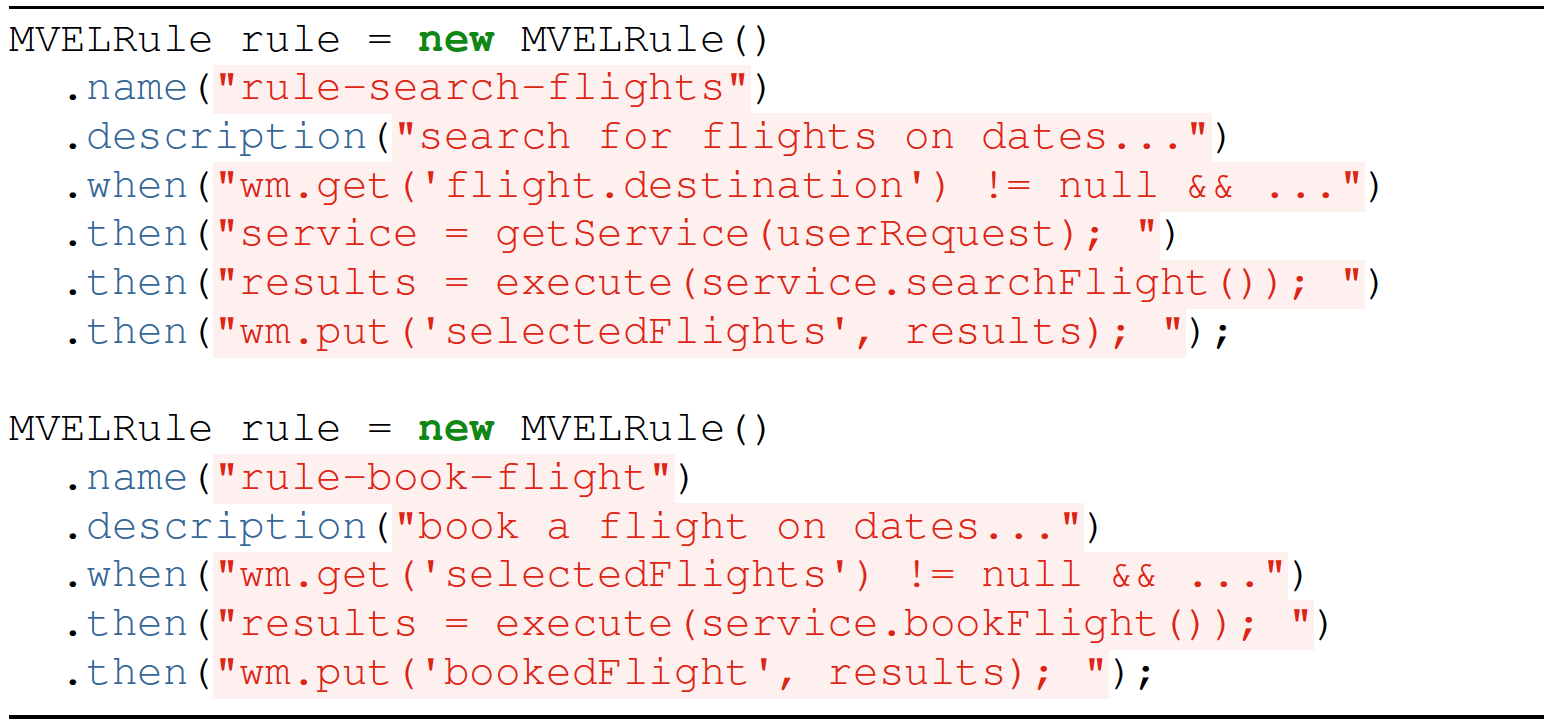}} 
    \vspace{-0.3cm}
    \caption{1. Excerpt for Compositional rule expressed in MVEL}
    \vspace{-0.3cm}
    \label{lst_mvel}
    \end{figure}
\else
\begin{listing}[h]
\begin{minted}
[
frame=lines,
framesep=1mm,
baselinestretch=1,
fontsize=\scriptsize
]{java}
MVELRule rule = new MVELRule()
  .name("rule-search-flights")
  .description("search for flights on dates...")
  .when("wm.get('flight.destination') != null && ...")
  .then("service = getService(userRequest); ")
  .then("results = execute(service.searchFlight()); ")
  .then("wm.put('selectedFlights', results); ");

MVELRule rule = new MVELRule()
  .name("rule-book-flight")
  .description("book a flight on dates...")
  .when("wm.get('selectedFlights') != null && ...")
  .then("results = execute(service.bookFlight()); ")
  .then("wm.put('bookedFlight', results); ");
\end{minted}
\caption{Excerpt for Compositional rule expressed in MVEL}
\label{lst_mvel}
\end{listing}
\fi

One of the main advantages of using NER along with MVEL rules is that unrestricted natural-language user requests can be easily transformed into programmatic compositional rules, for instance, when user says something like \emph{``look for flights to Paris on August 10 at 10:30am''} then NER will identify that substring ``August 10 at 10:30am'' refers to a single entity of type TIME, which becomes pretty convenient for further transformation into a Date Java object. Using the data disambiguation mechanism, this TIME entity can be mapped to argument ``departureDate'' on service method searchFlight, and finally, the MVEL rule mechanism can infer that the WM may have an object called ``flight'' that has an attribute named ``departureDate'', which can be evaluated as a precondition for service execution. The flexibility of this approach allows discovering and re-configuring types at runtime without linking to specific classes and objects at design-time.

\vspace{-0.2cm}
\subsection{Service Discovery, QoS-Aware Composition and Execution}
\label{sec_execution}

\textbf{\emph{Service Discovery:}} Once a composite service has been created (by linking abstract services through the use of compositional rules) then the service discovery mechanism searches for concrete services associated to each abstract service. To accomplish this task, service discovery uses the automated bundle discovery and registration mechanism provided by the OSGi Felix framework as described in Section~\ref{sec_toolchain}. 

\textbf{\emph{QoS-aware composition:}} After a set of concrete services is discovered per each abstract service, then it is necessary to select the most appropriate concrete service in terms of its QoS features. On early stages of development, developers may define a set of QoS features for each method per concrete service (as shown in \ifarxiv Figure \else listing \fi ~\ref{lst_yahoo}). Using a similar rule-based approach as described in the previous section, developers define a set of heuristics (rules) that are later validated using the rule engine. Every QoS has different triggering priorities, so for instance, \emph{battery consumption} has higher priority than \emph{connectivity} (since some services can still work locally even when there is no connectivity, but no services may work when battery is drained), which in turn has more priority than, let's say \emph{accuracy} (since two high-accuracy services may compete to be selected, however if they do not run locally but remotely and WiFi connection is disabled, then neither of them can be executed). Given a composite service (a sequence of linked abstract services resulting from the Service Coordination process described in section~\ref{sec_coordination}), and after validating the QoS features by firing the heuristics, a new set of concrete service method candidates are generated per each abstract service method, which is then passed to the Builder for service execution.

\textbf{\emph{Execution:}} Once a single concrete service associated to an abstract service has been selected, then it is executed. Execution can be performed either locally (e.g., if the service is running on a device) or remotely (e.g., in a server or the cloud). Required data to execute the service and service method is obtained from the WM, as explained above. After the service is executed, the current state of the composition is updated by adding the results of the execution to the WM. This process is repeated until no more abstract services are left.

\vspace{-0.3cm}
\section{Evaluation}
\label{sec_results}

We evaluated \middleware from three different perspectives in order to address the initial research questions: 
a precision/recall analysis to measure the performance of our system in real scenarios with users, a metric-based analysis to estimate the amount of effort (person/day) that could be minimized when using our approach vs. using a baseline approach~\footnote{Source code: \url{https://github.com/ojrlopez27/nl-service-composition}}, and a scalability test using different settings for \middleware.

\subsection{Performance: Recall, Precision and F1-Score}
\textbf{Setup:} for this experiment, we conducted a user study via Amazon Mechanical Turk where 20 users interacted with a chatbot~\cite{Zhao:2018}. We provided users 15 different services and 3 scenarios (plan a trip, plan a romantic dinner, and plan a party at home next weekend). Users were asked to describe what kind of requests (using unrestricted natural language) they would make to the chatbot for each of the three different scenarios. Conversations were logged and analyzed through a confusion matrix to determine the recall, precision and F1-score metrics. For this experiment, we used 2 pre-trained models, one uses 19.7B words from 700 dimensions trained on English tweets, and and the other uses 1.7B words from 700 dimensions trained on Wikipedia entries.

\textbf{Results:}
For the Actual Class, we defined two values: 1) YES: user's sentence is well structured, has meaning, can be understood, and should lead to the activation of a service and a specific method, and 2) NO: user's sentence is ambiguous, or out of context, or incomprehensible, or should not lead to the activation of a service (method is not available or does not exist). For the Predicted Class, we defined two values: 1) YES: \middleware has correctly identified the method and service OR if user sentence was ambiguous, then it should ask user to re-phrase the sentence, and 2) NO: \middleware selected a wrong service and method OR id did not ask user to re-phrase the sentence. Results are summarized in table~\ref{confusion}. Since we have an uneven class distribution, that is, false positives and false negatives are very different, then Accuracy metric is not of too much help, thus we need to rely on F1-score instead due to it computes the weighted average of Precision and Recall. Generally speaking, results demonstrate a good performance since the recall, precision an F1-Score are above 0.5. However, it is worth noting that these values might vary significantly from one experiment to the other since they rely on human judgment, which is bias-prone.

\vspace{-0.3cm}
\begin{table}[ht]
  \caption{Confusion Matrix for 3 scenarios and 20 participants}
  \vspace{-0.2cm}
  \label{confusion}
  \centering
\begin{tabular}{|ll|c|c|} 
\hline
& & \multicolumn{2}{l|}{ \textbf{Predicted class} }  \\ 
\cline{3-4}
& & \textbf{YES}  & \textbf{NO}                      \\ 
\hline
\multicolumn{1}{|l|}{\textbf{Actual Class} } & \textbf{YES}  & 341 & 109\\ 
\cline{2-4}
\multicolumn{1}{|l|}{} & \textbf{NO}   & 162 & 76 \\ 
\hline
\multicolumn{2}{|l|}{\textbf{Accuracy} } & \multicolumn{2}{c|}{0.60}\\ 
\hline
\multicolumn{2}{|l|}{\textbf{Recall} } & \multicolumn{2}{c|}{0.67} \\ 
\hline
\multicolumn{2}{|l|}{\textbf{Prediction} } & \multicolumn{2}{c|}{0.81} \\ 
\hline
\multicolumn{2}{|l|}{\textbf{F1-Score} } & \multicolumn{2}{c|}{0.74}\\
\hline
\end{tabular}
  \vspace{-0.5cm}
\end{table}

\subsection{Effort Estimation}
\textbf{Setup:} for this experiment, we estimated the total effort required to develop the \emph{``Plan-a-trip-to-Paris''} scenario (composed of 8 different services: \emph{FlightReservation}, \emph{HotelReservation}, \emph{Calendar}, \emph{Weather}, \emph{GroundTransportation}, \emph{Messaging}, \emph{LeisureActivities}, and \emph{Maps}) using 2 different approaches: \middleware vs. a Baseline service composition model (BSC) that uses WSDL templates for service descriptions, OWL-S for semantic matching\footnote{Matchmaker implementation: \url{https://bit.ly/2uZwznA}}, and BPEL4J for service coordination. As an effort estimation model we used COCOMO II (COnstructive COst MOdel II)~\cite{Boehm:2000}, a model that computes the effort (and cost) of a software project by fitting a regression formula based on a number of environmental factors related to systems engineering and historical data. 

\textbf{Results:} Using Lines of Code (LoC) metric as an input, COCOMO II computes software development effort as a function of program size and a set of 22 ``cost drivers'' that include subjective assessment of product, platform, personnel and project attributes; where each of them can be assigned a six-point scale rating (ranging from ``very low'' to ``extra high''). Therefore, the effort is calculated as: $E = a_i (KLoC)^{(b_i)} (EAF)$, where $E$ is the effort applied in person-months, $KLoC$ is the estimated number of thousands of LoC, and EAF is the Effort Adjustment Factor derived from the cost drivers. Constants $a_i$ and $b_i$ depend on the category of the system (organic, semi-detached, and embedded). We categorize our experiment as organic under the assumptions that the required development team is adequately small, the problem is well understood and the team members have a nominal experience regarding the problem. For an organic project, the values of $a_i$ and $b_i$ are 3.2 and 1.05 respectively~\cite{Boehm:2000}.
For the implementation of the experiment using \middleware we measured 2,562 LoC, while we got 3,267 LoC for BSC (including LoC for WSDL/OWL-S service descriptions). We assumed that most of the cost drivers remain the same for both approaches (i.e., nominal rating), the only cost driver that we consider may vary is ``Language and Toolset Experience (LTE)'' since developers may or may not have any experience on describing services and compositions using WSDL and OWL-S. For this reason, we considered 3 different scenarios: 1) \emph{Average-case scenario (A)}: developers have average experience with WSDL/OWL-S, so both \middleware and BSC have a nominal rating level for the LTE cost driver; 2) \emph{Worst-case scenario (W)}: developers have little or no experience with WSDL/OWL-S languages, so the LTE's rating level for BSC is ``very low''; and 3) \emph{Best-case scenario (B)}: developers are experts on WSDL/OWL-S so LTE's rating level for BSC is ``extra high''. In scenarios 2 and 3, \middleware's LTE remains nominal since there are no special developer's skills required to generate the annotations. Also, using COCOMO II we estimated the software project schedule (months) and cost (dollars). Finally, we calculated an improvement rate of \middleware over BSC according to each scenario (Rate-A, Rate-W, and Rate-B, respectively). 
The results on Table~\ref{effort}\footnote{Using COCOMO II calculator: \url{http://csse.usc.edu/tools/COCOMOII.php}} show that, in general, \middleware significantly reduces the amount of total effort in comparison with the 3 configurations of BSC (ranging from a 8.79\% improvement rate for the best-case scenario to a 36.15\% improvement rate for the worst-case scenario). \middleware promises to reduce the cost of the project in similar proportions, while reducing the project schedule in approximately a third of the improvement rates for effort and cost estimations (i.e., 2.63\% - 12.94\%). The schedule estimation was not reduced in the same proportions as effort and cost due to it uses different calibrate scale factors that reflect changing requirements, CI/CD, and other variables that may affect the estimation.

\begin{table}[ht]
\centering
\vspace{-0.2cm}
\caption{Effort Estimation for \middleware vs. a Baseline Approach (BSC)}
\vspace{-0.2cm}
\label{effort}
\resizebox{\columnwidth}{!}{%
\begin{tabular}{|l|r|r|r|r|r|r|r|} 
\hline
\textbf{Metric}                & \textbf{NLSC-A} & \textbf{BSC-W} & \textbf{BSC-A} & \textbf{BSC-B} & \textbf{Rate-W} & \textbf{Rate-A} & \textbf{Rate-B}  \\ 
\hline
\textbf{LoC}             & 2,562             & 3,267                 & 3,267                 & 3,367                 & 21.58\%           & 21.58\%           & 21.58\%            \\ 
\hline
\textbf{Effort}   &  8.3  & 13.0 & 10.8 & 9.1 & 36.15\% & 23.15\% & 8.79\%   \\\hline
\textbf{Schedule}    & 7.4 & 8.5  & 8.1 & 7.6 & 12.94\% & 8.64\% & 2.63\%   \\\hline
\textbf{Cost}    &  \$33K  & \$52K  & \$43K & \$36K & 36.54\% & 23.26\% & 8.33\%  \\\hline
\end{tabular}
}
\vspace{-0.2cm}
\end{table}

It is worth noting that, despite the fact that \middleware reduces the number of LoC in relation to the BSC implementation at a rate of 21.58\%, the estimation produces lower improvement rates for the best-case scenario and higher improvement rates for the worst-case scenario. This means that, even in comparison with the BSC best-case scenario where the development team counts on the participation of highly skilled service developers (i.e., developers that are experts on conventional service tools, languages and technologies such as WSDL, OWL-S, BPEL, etc.), who can easily create software service-based solutions and prototypes; \middleware can still reduce the amount of effort, cost and schedule for the software project development. Of course, this effort improvement is more evident in the worst-case scenario, making possible that developers with no service-oriented training can easily generate software solutions from simple prototypes to comprehensive enterprise solutions.

\subsection{Scalability Test}
\vspace{-0.1cm}
\textbf{Setup:} the purpose of this experiment is to find out how much \middleware's performance is improved or degraded when the system is scaled to thousands of services. To this purpose, we measured the system's response time (ms) for the plan-to-trip-scenario when using either 5, 50, 500, or 5,000 service descriptions in the corpora. We used service descriptions from a web service dataset\footnote{QWS dataset: \url{http://www.uoguelph.ca/~qmahmoud/qws/index.html}}. The sent2vec module was trained using two models: one model with 19.7 billion sentences from Tweeter entries, and another model with 1.9 billion sentences from Wikipedia entries. Then, we ran the experiment 10 times per setting and computed the harmonic mean (Harm.) of the response times (the harmonic mean mitigates the impact of large outliers and aggravates the impact of small ones). We estimated the standard deviation (StdDev), the time response per service (TPS) calculated as (Harm / \# services), and an improvement/degradation rate calculated as $1- (TPS_{1} / TPS_{-1}) \%$. The experiment was run on a MacBook Pro Intel Core i7, 1.7 GHz, 8 GB, 1600 MHz DDR3.

\textbf{Results:} the results of this experiment are shown in table~\ref{scalability}. Generally speaking, the system's performance is not affected when progressively scaled the system up to thousands of services, actually, one can observe an improvement rate that decreases inversely proportional the number of services. In particular, we observed that the time response per service decreased from 5 to 50 to 500 to 5,000 services, though not in the same proportions. For instance, for the Wiki model, when using only 5 services the composition took 0.022 ms per service, while taking 0.005 ms per service when using 50 services, which corresponds to a 77.21\% of improvement. The reason why the performance of the system is not affected when increasing the number of services is that the composition model keeps some information in memory which can be reused while constructing the composite service, for instance, when resolving and mapping the service method ``searchFlights'', arguments ``fromOrigin'', ``toDestination'', ``departureDate'', ``arrivalDate'', etc. are stored in the working memory. Later on, when resolving service method ``bookFlight'', almost all the arguments can be easily retrieved from the working memory without requiring to perform data-type disambiguation and firing compositional rules. Likewise, service method ``searchHotels'' and ``bookHotel'' can retrieve information from the working memory such as ``toDestination'', ``departureDate'', and ``arrivalDate''. Regarding the two pre-trained models (Wikipedia and Twitter) we observed that using Twitter took more time for the system to perform the semantic matching based on cosine similarity (mostly due to the considerable difference in size of both models), however, \middleware increased its accuracy in almost 28\%, which leads to the conclusion that there should be a trade-off between the accuracy of the mappings between the user request and the service description, and the system's response time (performance) which will increase when the model is bigger in size.

\begin{table}[ht]
\centering
\vspace{-0.2cm}
\caption{Scalability}
\vspace{-0.2cm}
\label{scalability}
\resizebox{\columnwidth}{!}{%
\begin{tabular}{|r|r|r|r|r|r|r|r|r|} 
\hline
                   & \multicolumn{4}{c|}{ \textbf{Wiki Model }}                               & \multicolumn{4}{c|}{ \textbf{Twitter Model }}                            \\ 
\hline
\textbf{Services} & \textbf{Harm.} & \textbf{StdDev} & \textbf{TPS}  & \textbf{Rate} & \textbf{Harm.}  & \textbf{StdDev} & \textbf{TPS}  & \textbf{Rate}  \\ 
\hline
5                  & 0.12               & 2.60              & 0.022          & 0\%            & 0.16               & 0.59              & 0.032         & 0\%             \\ 
\hline
50                 & 0.28               & 0.34              & 0.005          & 77.21\%        & 0.39               & 0.98              & 0.007         & 75.07\%         \\ 
\hline
500                & 1.39               & 1.54              & 0.002          & 51.79\%        & 1.64               & 1.38              & 0.003         & 58.85\%         \\ 
\hline
5,000              & 7.60               & 2.13              & 0.001          & 45.37\%        & 8.05               & 1.26              & 0.001         & 50.98\%         \\
\hline
\end{tabular}
}
\end{table}

\vspace{-0.3cm}
\section{Related Work}
\label{sec_related_work}

Users interact instinctively with the system in an easily expressible natural language and thus expect the system to identify the set of services that are required to achieve the user's goal. In our study, we review natural language-based approaches for dynamic service composition. If we consider an user’s natural language description at one end of the problem and services at the other end, then, we find that existing literature can be broadly categorized as approaches that a) apply restrictions on how the user expresses the goal using sentence templates and/or user utterances and then use structured parsing techniques to parse the sentences against service descriptions \cite{DBLP:journals/jsw/BoscaCVM06,6384197}; b) construct semantic graphs that represent the service description \cite{1546104} \cite{Romero:2019} \cite{Romero:concise:2019} such that those could be matched with the natural language descriptions using a lexical database such as WordNet, that groups words based on their meanings, to calculate a conceptual distance metric between concepts \cite{4801832} \cite{5283917}; and c) match partially-observable natural language description using semantic web services such as OWL-S\cite{10.1007/978-3-642-17358-5_54} \cite{CHARIF200633}. Categorical limitations of existing approaches include, (i) complex linguistic processing that employs several NLP techniques: structured parsing, extracting parts-of-speech tokens, stop-word removal, spell-checking, stemming, and text segmentation, (ii) inclusion of lexical databases such as WordNet or domain-specific ontologies that represents domain lexicons, and (iii) a weaker concept representation and similarity score for semantic matching that does not account for sentence context. 
To overcome the above limitations, in our work, we (a) allow users to express their sentences template-free and use their natural language description without complex linguistic processing by aligning it with service descriptions using Sentence embeddings, (b) avoid the need for lexical databases and ontologies by relying on the automatically extracted corpus of service descriptions which would otherwise be provided by service developers as code comments on services; this reduces the need to construct semantic graphs of concepts and domain-specific ontologies, and (c) use a stronger representation of words, concepts and natural language sentences that account for word usage in context to user’s sentence by applying a state-of-the-art pre-trained semantic representation model of English language.

\vspace{-0.1cm}
\section{Conclusions and Future Work}
\label{sec_conclusions}
In this paper we have presented \middleware, a Dynamic Service Composition Middleware based on unrestricted Natural Language service descriptions. Using our approach, we have demonstrated that total effort (in terms of person/day) for service composition and service integration can be dramatically reduced up to 36\% thanks to we eliminated the Translation phase proposed by the Service Composition Middleware (SCM) model and substituted it by an intuitive mechanism for service description, discovery, and retrieval. We also demonstrated that Service Composition using Sentence Embeddings and Named Entity Recognition techniques alleviate the burdensome task of writing boilerplate code, strictly defining well-defined hard-typed interfaces, validating ontology models and representations, and creating ad-hoc semantic reasoning mechanisms for service matching. 
Also, we estimated that the cost-overhead of using extra comments in the code is minimal since developers would only have to learn a reduced number of code annotations (i.e., @Description, @ArgDesc, and QoS annotations). These annotations are easy to document since they do not require any particular structure (they are plain natural language-based descriptions) and resemble the structure of conventional Java annotations and Java documentation. One of the limitations of our approach is that it is only oriented to Java (for now), however, we are planning to make it available to other programming languages by removing the need of Java annotations and allowing developers to write their service descriptions on plain text files.

\textbf{\emph{Future Work:}} we plan to improve the precision of our model by training custom service description models in addition to common-sense pre-trained models as Wikipedia or Twitter entries. Also, we plan to extend our approach so it can discover third-party services published in well-known public repositories such as ProgrammableWeb.com and GitHub. 

\textbf{\emph{Discussion:}} data-driven ML and NLP approaches raise several open questions including learning with limited data. For instance, a) learning QoS-aware models that introduce model sparsity, b) inferring custom entities using reinforcement and online learning, with initial disambiguations by user, to improve service matches, c) learning context-sensitive models with working memory for better entity resolution, and d) one-shot learning from descriptions for service disambiguation.

\hypersetup{
    urlcolor=black
}

\vspace{-0.2cm}
\bibliographystyle{IEEEtranS}
\bibliography{main}

\end{document}